\documentclass[prb,reprint]{revtex4-1}
\usepackage[latin1]{inputenc}
\usepackage{graphicx}
\usepackage{amssymb}
\usepackage{amsmath}
\usepackage{xspace}
\usepackage{dcolumn}
\usepackage{bm}
\usepackage{color}
\usepackage{float}
\usepackage[extra]{tipa}

\setcitestyle{square,numbers}

\newfont{\tensy}{cmsy10}

\newcommand{\ie}[0]{i.e.\@\xspace}
\newcommand{\eg}[0]{e.g.\@\xspace}

\newcommand{\rmi}{\text{i}}
\newcommand{\UP}[0]{\uparrow}
\newcommand{\DO}[0]{\downarrow}

\newcommand{\oP}{\hat{P}}
\newcommand{\oQ}{\hat{Q}}

\newcommand{\on}{\hat{n}}

\newcommand{\rmd}{\text{d}}

\newcommand{\bit}{\begin{itemize}}
\newcommand{\eit}{\end{itemize}}

\newcommand{\oh}{\mbox{$\frac{1}{2}$}}

\newcommand{\om}[0]{\omega}

\newcommand{\kB}{k_\text{B}}
\newcommand{\nag}{{\phantom{\dag}}}

\newcommand{\las}[0]{\langle}
\newcommand{\ras}[0]{\rangle}

\begin{document}

%%%%%%%%%%%%%%%%%%%%%%%%%%%%%%%%%%%%%%%%%%%%%%%%%%%%%%%%%%%%%%%%%%%%%%%%%%%%%
%%%%%%%%%%%%%%%%%%%%% TITLE & ABSTRACT %%%%%%%%%%%%%%%%%%%%%%%
%%%%%%%%%%%%%%%%%%%%%%%%%%%%%%%%%%%%%%%%%%%%%%%%%%%%%%%%%%%%%%%%%%%%%%%%%%%%%

\title{Two-dimensional Holstein-Hubbard model:\\ Critical temperature, Ising universality, and bipolaron liquid}

\author{Manuel Weber}

\affiliation{\mbox{Institut f\"ur Theoretische Physik und Astrophysik,
    Universit\"at W\"urzburg, 97074 W\"urzburg, Germany}}

\author{Martin Hohenadler}

\affiliation{\mbox{Institut f\"ur Theoretische Physik und Astrophysik,
    Universit\"at W\"urzburg, 97074 W\"urzburg, Germany}}

\begin{abstract} 
  The two-dimensional Holstein-Hubbard model is studied by means of continuous-time
  quantum Monte Carlo simulations. Using renormalization-group-invariant
  correlation ratios and finite-size extrapolation, the critical temperature of the
  charge-density-wave transition is determined as a function of coupling
  strength, phonon frequency, and Hubbard repulsion. The phase
  transition is demonstrated to be in the  universality class of the
  two-dimensional Ising model and detectable via the fidelity susceptibility.
  The structure of the ground-state phase diagram and the possibility of a bipolaronic metal with
  a single-particle gap above $T_c$ are explored.
\end{abstract}

\date{\today}

\maketitle

\section{Introduction}\label{sec:introduction}

Phase transitions in two-dimensional (2D) fermionic
systems are a central topic of theoretical and experimental condensed matter
physics. Correlated quasi-2D materials with rich phase diagrams include
high-temperature superconductors \cite{RevModPhys.78.17} and transition-metal
dichalcogenides \cite{manzeli20172d}. Dirac fermions in two dimensions can be
investigated in graphene \cite{Neto_rev}. 
Strongly correlated 2D fermions exhibit exotic phases
\cite{RevModPhys.89.025003} and phase transitions
\cite{Senthil04_2}, and can support long-range order
at $T>0$ \cite{PhysRevLett.17.1133}. While magnetism originates
from short-range Coulomb repulsion, the main mechanism behind the numerous
charge-density-wave (CDW) phases found experimentally is electron-phonon
coupling. In addition to polaron effects, the latter leads to
a phonon-mediated, retarded electron-electron interaction and an intricate
interplay of spin, charge, and lattice fluctuations. 

Quantum Monte Carlo (QMC) simulations are a key tool to investigate 
correlated 2D quantum systems. Although simulations
are significantly harder for fermions than for spins or bosons, QMC methods have been very
successfully applied to fermionic models. However, whereas the phase diagram
and critical behavior of, \eg, the 2D honeycomb Hubbard model is known in detail
\cite{Sorella12,Assaad13,Toldin14,Otsuka16}, the same is not true even for
the simplest Holstein molecular-crystal model of electron-phonon
interaction. Most notably, simulations with phonons are often severely restricted
by long autocorrelation times also away from critical points \cite{Hohenadler2008}. 
Currently, reliable critical temperatures, convincing analysis of critical
behavior, and the ground-state phase diagram remain key open problems. In fact, even the
simpler 1D case had until recently been discussed controversially
\cite{MHHF2017}, with earlier claims of dominant pairing correlations refuted
by direct calculations of the correlation functions and traced back to spin
gap formation \cite{PhysRevB.92.245132}.

Here, we use large-scale continuous-time QMC simulations to investigate
the CDW transition in the 2D Holstein-Hubbard model. Although the latter has
been extensively studied in the past, important open questions remain.
At strong coupling and half-filling, the ground state is either a CDW
insulator or an antiferromagnetic Mott insulator. Recent variational QMC studies
\cite{ohgoe2017competitions,1709.00278}  argue in
favor of a third phase (metallic or superconducting), seemingly in
contradiction with theoretical arguments based on weak-coupling
instabilities of the Fermi liquid \cite{PhysRevLett.56.2732,PhysRevB.42.2416}. We use finite-size
scaling to determine $T_c$ of the CDW transition, show that the latter can
also be detected by the fidelity susceptibility, and provide evidence for its
Ising critical behavior. Moreover, we present arguments and data for the
existence of a metallic bipolaron phase at $T>T_c$ and address the possibility of a
metallic or a superconducting ground state.

The paper is organized as follows. Section~\ref{sec:model} introduces the
relevant models, Sec.~\ref{sec:methods} gives a brief review of the numerical
methods, Sec.~\ref{sec:results} discusses the results, and Sec.~\ref{sec:conclusions} 
contains our conclusions.

\section{Models}\label{sec:model}

The Holstein-Hubbard Hamiltonian \cite{HOLSTEIN1959325} reads
\begin{align}\label{eq:model}\nonumber
  \hat{H}
  =
    &-t \sum_{\las i,j\ras \sigma} \hat{c}^\dag_{i\sigma} \hat{c}^\nag_{j\sigma} 
    +
    \sum_{i}
    \left[
    \mbox{$\frac{1}{2M}$} \oP^2_{i}
    +
    \mbox{$\frac{K}{2}$} \oQ_{i}^2
    \right]
  \\
    &-
      g
    \sum_{i} \hat{Q}_{i} 
     \hat{\rho}_i
    + U 
      \sum_i (\on_{i\UP}-\oh) (\on_{i\DO}-\oh)
    \,.
\end{align}
The first two terms describe free electrons and free phonons, respectively.
Here, $\hat{c}^\dag_{i\sigma}$ creates an electron with spin $\sigma$ at lattice
site $i$ and electrons hop with amplitude $t$ between nearest-neighbor
sites on a square lattice. The phonons are of the Einstein type with frequency
$\omega_0=\sqrt{K/M}$; their displacements $\hat{Q}_i$ couple to local
fluctuations $\hat{\rho}_i=\on_i-1$ of the electron occupation $\on_{i} =
\sum_\sigma \on_{i\sigma}$ where $\on_{i\sigma} = \hat{c}^\dag_{i\sigma}\hat{c}^\nag_{i\sigma}$. The
last term describes a Hubbard onsite repulsion of strength $U$.
We simulated $L\times L$ lattices with periodic boundary conditions at
half-filling  ($\las\on_{i}\ras=1$, chemical potential $\mu=0$). A
useful dimensionless coupling parameter is $\lambda=g^2/(W K)$ with the free
bandwidth $W=8t$. We set $\hbar$, $\kB$, and the lattice constant to one and
use $t$ as the energy unit.

For $U=0$, Eq.~(\ref{eq:model}) reduces to the Holstein model. Its
relative simplicity has motivated numerous  QMC investigations of CDW
formation and superconductivity \cite{%
PhysRevB.40.197,PhysRevB.42.2416,PhysRevB.42.4143,PhysRevLett.66.778,PhysRevB.43.10413,PhysRevB.46.271,PhysRevB.48.7643,PhysRevB.48.16011,PhysRevB.55.3803}. Equation~(\ref{eq:model})
with $g=0$ corresponds to the repulsive Hubbard model on the square lattice. At half-filling, the ground state of the latter is an
antiferromagnetic Mott insulator for any $U>0$ \cite{Hirsch89}. However, in contrast to 
CDW order, antiferromagnetism is restricted to $T=0$ in two dimensions by the
Mermin-Wagner theorem \cite{PhysRevLett.17.1133}.
The full Holstein-Hubbard Hamiltonian~(\ref{eq:model}) captures the
competition between Mott and CDW ground states
\cite{PhysRevB.52.4806,PhysRevLett.75.2570,PhysRevB.75.014503,PhysRevB.92.195102,PhysRevLett.109.246404,PhysRevB.87.235133,ohgoe2017competitions}.
Whereas early work unanimously agreed on the absence of a disordered
or a superconducting ground state at half-filling, such a phase has recently
been advocated by numerical results \cite{ohgoe2017competitions,1709.00278}. 

\begin{figure}[t]
  \includegraphics[width=0.45\textwidth]{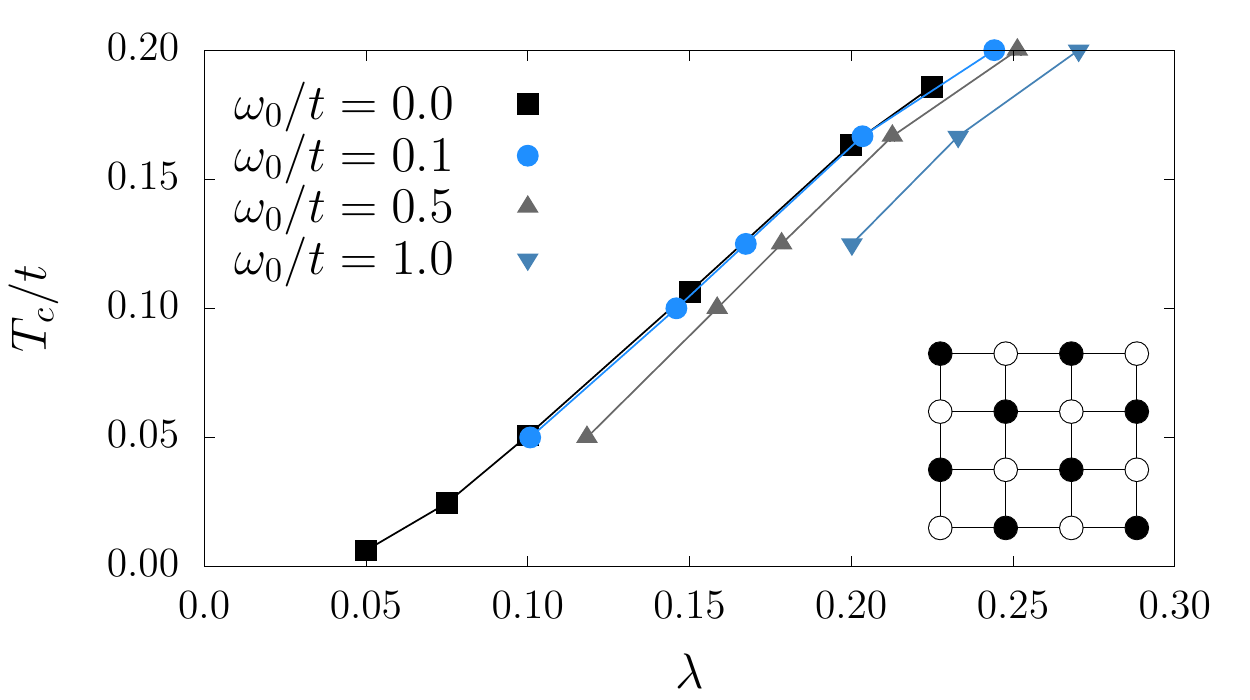}
  \caption{\label{fig:phasediagram} Critical temperature of the Holstein
    model ($U=0$) from finite-size scaling.  Here and in subsequent figures, lines are guides
    to the eye. Statistical errors are smaller than the symbols, see text.
    The inset illustrates the CDW order at $T<T_c$ for $L=4$, with filled
    (open) symbols representing occupied (empty) sites.}
\end{figure}

Because it is sufficient to address many of the open questions of interest,
we will mainly consider the case $U=0$. However, selected results for
the impact of the Hubbard repulsion will also be reported. For
Eq.~(\ref{eq:model}) with $U=0$, mean-field theory (exact for $\omega_0=0$ and
$T=0$) predicts a CDW ground state with a checkerboard pattern for
the lattice displacements and the charge density [ordering vector ${\bm
  Q}=(\pi,\pi)$, see inset of Fig.~\ref{fig:phasediagram}]  
at half-filling \cite{PhysRevB.40.197,PhysRevB.42.2416,PhysRevLett.66.778}.
Here, we systematically explore the impact of quantum and thermal fluctuations.

An important limiting case is the antiadiabatic limit $\omega_0\to\infty$, in
which the Holstein-Hubbard model maps to a Hubbard model with Hamiltonian
\begin{align}\label{eq:model2}%\nonumber
  \hat{H}
  &=
    -t \sum_{\las i,j\ras \sigma} \hat{c}^\dag_{i\sigma} \hat{c}^\nag_{j\sigma} 
    +
    U_\infty    
    \sum_{i} (\on_{i\UP}-\oh) (\on_{i\DO}-\oh)
\end{align}
and effective interaction $U_\infty=U-\lambda W$. For $U=0$, interactions are
purely attractive and give rise to coexisting CDW and superconducting order
for any $\lambda>0$ at $T=0$. However, at half-filling, this order is minimal
in the sense that $T_c=0$ \cite{Hirsch85}, which is related to a perfect
degeneracy of CDW and pairing correlations and an associated continuous SO(3)
order parameter for which the Mermin-Wagner theorem applies \cite{PhysRevLett.17.1133}.

\section{Methods}\label{sec:methods}

Extending previous applications to 1D electron-phonon models \cite{Ho.As.Fe.12,PhysRevB87.075149,PhysRevLett.117.206404}, we use the
continuous-time interaction expansion (CT-INT) method \cite{Rubtsov05}. To
this end, we express the partition function as a functional integral
\begin{align} \label{partitionfunction}
 Z = \int \mathcal{D}(\bar{c},c) \ e^{-S_0\left[\bar{c},c\right]-S_1\left[\bar{c},c\right]} \int \mathcal{D}(\bar{b},b) \ e^{-S_\text{ep}\left[\bar{c},c,\bar{b},b\right]}
\end{align}
using coherent states.
Splitting the action into the free-fermion part $S_0$, the Hubbard interaction $S_1$, 
and the remainder $S_\text{ep}$ that contains the free-phonon contribution
and the electron-phonon coupling, the phonons are integrated out analytically
to arrive at a fermionic model with both an instantaneous Hubbard interaction ($S_1$)
and a retarded, phonon-mediated interaction ($S_2$) \cite{Assaad07}. This model can
be simulated by the CT-INT method by sampling both types of vertices
\cite{Assaad07} to stochastically sum the weak-coupling Dyson expansion
\cite{Rubtsov05} around $S_0$. Because the latter converges for fermionic systems in
a finite spacetime volume, CT-INT is exact apart from statistical errors.
Technical reviews can be found in Refs.~\cite{Gull_rev,Assaad14_rev}.

In contrast to the determinant QMC (DetQMC) method \cite{Blankenbecler81}
used in almost all previous works on Holstein-Hubbard-type models, CT-INT
has significantly smaller autocorrelation times \cite{Hohenadler2008}.
CT-INT simulation times scale as ${O}(n^3)$, where $n$
[$\approx {O}(\lambda\beta L^2)$ for $U=0$] is the average expansion
order and $\beta=1/T$. Although DetQMC formally has a better
${O}(\beta L^6)$ scaling, CT-INT benefits from
reduced expansion orders at weak coupling and seems to outperform DetQMC for
most parameters considered despite being limited for $\omega_0\gtrsim t$ by a
sign problem. Whereas even the noninteracting case is challenging for DetQMC, CT-INT
trivially gives exact results for $\lambda=0$ and can in principle
simulate the entire range of phonon frequencies, including the
experimentally important adiabatic regime $\omega_0<t$. We used up to 5000
single-vertex updates and 8 Ising spin flips per sweep.
The classical case $\omega_0=0$ was simulated using the method
of Ref.~\cite{1996MPLB...10..467M} combined with parallel tempering \cite{doi:10.1143/JPSJ.65.1604}.

\section{Results}\label{sec:results}

Since the effect of electron-electron repulsion on a half-filled square
lattice---namely, an antiferromagnetic Mott state at $T=0$---is well
understood \cite{Hirsch89} the focus of our work will be the electron-phonon
interaction, \ie, Eq.~(\ref{eq:model}) with $U=0$. Coulomb interactions of
the Hubbard or even long-range type can be simulated with unbiased QMC
methods on systems large enough to extract critical exponents \cite{Sorella12,Assaad13,Toldin14,Otsuka16}.
In contrast, the electron-phonon interaction is significantly more
challenging to describe due to the resulting, retarded electron-electron
interaction. This is true both for QMC methods but also for, \eg, the
functional renormalization group \cite{PhysRevB.75.014503,PhysRevB.92.195102}. Consequently, even fundamental aspects such
as the existence of a nonzero critical value for the CDW transition are still
under debate. From the 1D Holstein-Hubbard model, it is known that the phases
at $U=0$ (Luther-Emery liquid and CDW insulator) are stable against a nonzero
Hubbard repulsion so that they and the phase transition between them can be
fully understood in the simpler Holstein model \cite{MHHF2017}. In particular, 
the metallic phase arises from quantum lattice fluctuations 
rather than from a competing Hubbard interaction \cite{MHHF2017}. In the 2D case, recent
predictions of an extended metallic region suggest that the latter is largest
at $U=0$ \cite{ohgoe2017competitions,1709.00278}. A nonzero but sufficiently
small Hubbard repulsion merely shifts the critical value for the CDW transition
\cite{PhysRevB.52.4806,PhysRevLett.75.2570,PhysRevB.75.014503,PhysRevB.92.195102,PhysRevLett.109.246404,PhysRevB.87.235133,ohgoe2017competitions}.
Moreover, because the long-range antiferromagnetic order is restricted to
$T=0$, the possible phases at $T>0$ (the focus of this work) remain the same. 

\begin{figure}[b]
  \includegraphics[width=0.45\textwidth]{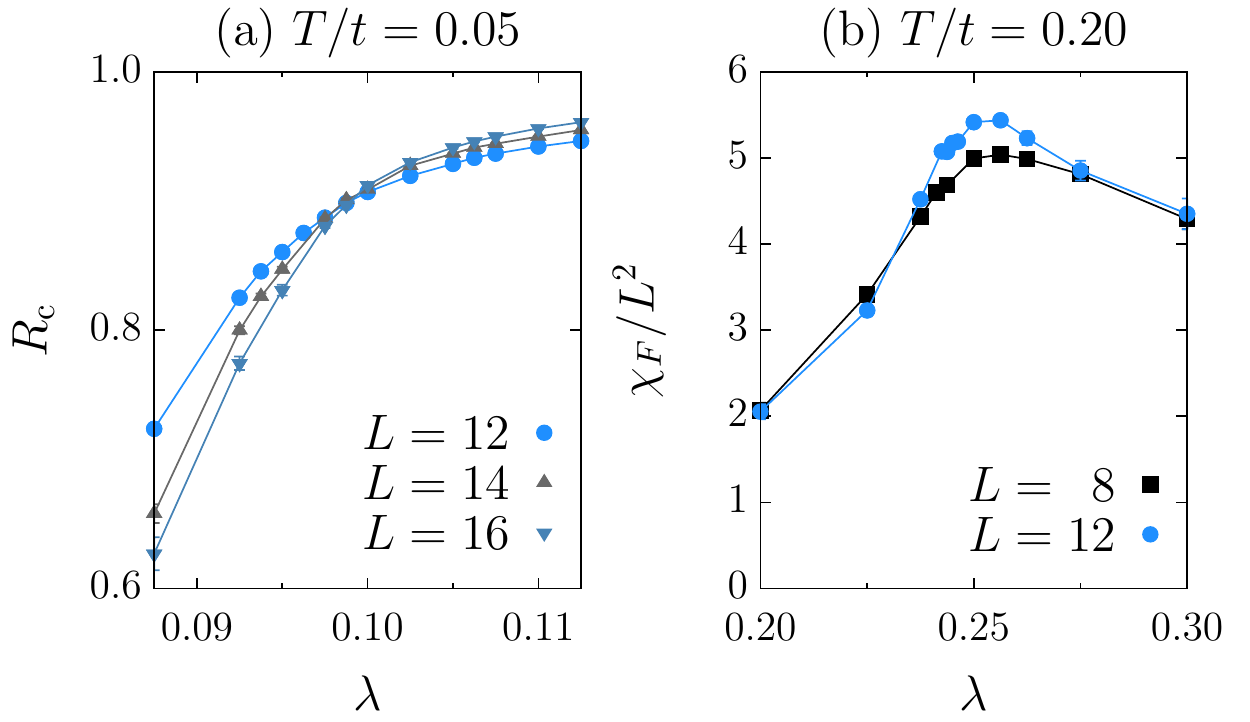}  
  \caption{\label{fig:criticalpoint} (a) Determination of the critical value $\lambda_c$ 
    from (a) the crossing of the correlation ratios $R_\text{c}$ for
    different system sizes $L$ and (b) the maximum in the fidelity
    susceptibility. Here, $\omega_0/t=0.1$, $U=0$, and (a) $T/t=0.05$, (b)
    $T/t=0.2$. 
  }
\end{figure}

\subsection{Critical values}\label{sec:results:Tc}

\begin{figure}[t]
  \includegraphics[width=0.45\textwidth]{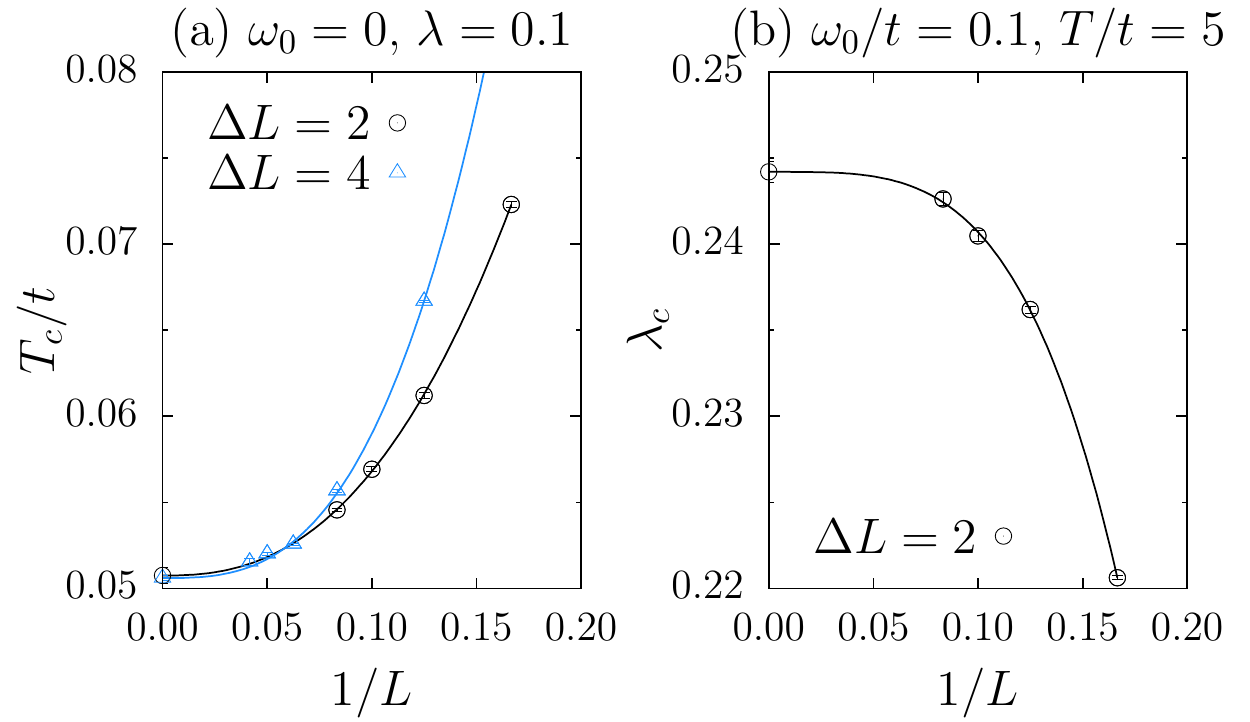}
  \caption{\label{fig:extrapolation}  
    Finite-size extrapolation of the crossing points of $R_\text{c}(L)$,
    $R_\text{c}(L-\Delta L)$ using the fit function~(\ref{eq:fitfunction}). 
    Here, (a) $\omega_0=0$, $\lambda=0.1$, $T_c=0.0506(1)$ and (b)
    $\omega_0/t=0.1$, $T/t = 0.2$, $\lambda_c=0.244(1)$.
  }
\end{figure}

To obtain the critical values shown in
Fig.~\ref{fig:phasediagram}, we calculated the correlation ratio
\cite{Binder1981}
\begin{equation}\label{eq:rcdw}
  R_\text{c} =
  1-\frac{S_\text{c}(\bm{Q}-\delta{\bm q})}{S_\text{c}(\bm{Q})}
\end{equation}
(with $|\delta{\bm q}|=2\pi/L$) from the charge structure factor
\begin{equation}\label{eq:scdw}
S_\text{c}({\bm q}) = \frac{1}{L^2}\sum_{ij}
e^{\rmi(\bm{r}_i-\bm{r}_j)\cdot\bm{q}}\,\las \on_i \on_j\ras
\end{equation}
either at fixed $\lambda$ or at fixed $T$. Here, ${\bm Q}=(\pi,\pi)$. 
By definition, a divergence of $S_\text{c}(\bm{Q})$ with $L$ in the CDW phase
implies $R_\text{c}\to 1$ for $L\to\infty$, whereas $R_\text{c}\to 0$ in
the absence of long-range CDW order. Moreover, because $R_\text{c}$ is
a renormalization group invariant \cite{Binder1981}, the critical point can be
estimated from the crossing of curves for different $L$, as illustrated in
Fig.~\ref{fig:criticalpoint}(a) for $\omega_0/t=0.1$ and $T/t=0.05$. 
While the correlation ratio~(\ref{eq:rcdw}) is expected to exhibit smaller
finite-size corrections than the structure factor~(\ref{eq:scdw}), a shift
of consecutive crossing points is observed on the accessible system sizes,
making it necessary to extrapolate to
$L=\infty$. To this end, we used a fit function
\begin{equation}\label{eq:fitfunction}
f(L) = a + b L^c\,.
\end{equation}
Examples for such extrapolations are shown for $\om_0/t=0$ in Fig.~\ref{fig:extrapolation}(a)
and for $\om_0/t=0.1$ in Fig.~\ref{fig:extrapolation}(b). For classical
phonons, we can access significantly larger system sizes up to $L=28$. The
points in Fig.~\ref{fig:extrapolation}(a) correspond to crossing points of
$R_\text{c}$ for $L$, $L-2$ (\ie, $\Delta L=2$) and $L$, $L-4$ ($\Delta
L=4$), respectively. Fitting to Eq.~(\ref{eq:fitfunction}), these two choices
yield identical results for $T_c$ within error bars. The errors take into
account the statistical errors of the QMC results as well as the errors in
determining the crossing points using parabolic fits (obtained from a
bootstrap analysis) and extrapolating to $L=\infty$. They are smaller than
the symbol size in Fig.~\ref{fig:phasediagram} but naturally do not capture
possible variations due to the choice of fit function or observable.  
For quantum phonons, we systematically used $L=4,6,8,10,12$ and hence $\Delta L=2$, as
illustrated in Fig.~\ref{fig:extrapolation}(b). A similar extrapolation gives
$\lambda_c=0.101(1)$ for the parameters of Fig.~\ref{fig:criticalpoint}(a).

The phase transition can also be detected using the fidelity susceptibility
$\chi_{F}$ \cite{2008arXiv0811.3127G}, an unbiased diagnostic to
detect critical points without any knowledge about the order parameter.
It essentially relies on calculating the overlap of the ground states of (in
the present case) Holstein 
Hamiltonians with couplings $\lambda$ and $\lambda+\delta\lambda$. A
finite-temperature generalization has been given in Refs.~\cite{PhysRevE.76.022101,PhysRevLett.103.170501, PhysRevB.81.064418}
and CT-INT estimators in Refs.~\cite{PhysRevX.5.031007,PhysRevB.94.245138}.
Although these estimators have rather large statistical errors at low
temperatures, $\chi_{F}/L^2$ for $T/t=0.20$ in
Fig.~\ref{fig:criticalpoint}(b) shows the expected peak at a position that is
consistent with Fig.~\ref{fig:phasediagram} and
$\lambda_c=0.244(1)$ from Fig.~\ref{fig:extrapolation}(b).

Figure~\ref{fig:phasediagram} shows  $T_c(\lambda)$ for
different $\omega_0$, covering the entire adiabatic regime $0\leq
\omega_0 \leq t$. The mean-field result $T_c\sim e^{-1/\sqrt{\lambda}}$ for
the 2D Holstein model---compared to $T_c\sim e^{-1/\lambda}$ in dynamical mean-field
theory (DMFT) \cite{PhysRevB.63.115114}---is expected to overestimate
$T_c$ even at $\omega_0=0$ and does not capture the expected maximum at
$\lambda<\infty$ \cite{PhysRevB.63.115114}. The latter is outside
the range of couplings considered here. Quantum lattice fluctuations
suppress $T_c$ at a given $\lambda$. For $\omega_0/t=0.1$, $T_c$ shows only
minor deviations from the result for classical phonons, whereas for larger $\omega_0$ 
quantum fluctuation effects are clearly visible over the entire parameter range shown.
 The systematic suppression of $T_c$ with
increasing $\omega_0$ is perfectly consistent with the fact that $T_c=0$ for the
attractive Hubbard model \cite{Hirsch85}, to which the Holstein model maps in the limit
$\omega_0\to\infty$ \cite{Hirsch83a}. This connection and a possible metallic phase at low temperatures
as a result of quantum fluctuations will be discussed below. At $T>0$, a
metallic region is naturally expected in the phase diagram of the 2D
Holstein-Hubbard model because  the antiferromagnetic Mott state arising from
the Hubbard interaction is confined to $T=0$. In contrast to previous  DMFT results
\cite{PhysRevB.63.115114}, the critical temperatures in
Fig.~\ref{fig:phasediagram} were obtained by taking into account all (spatial
and temporal) fluctuations on the square lattice. 

\begin{figure}[t]
  \includegraphics[width=0.45\textwidth]{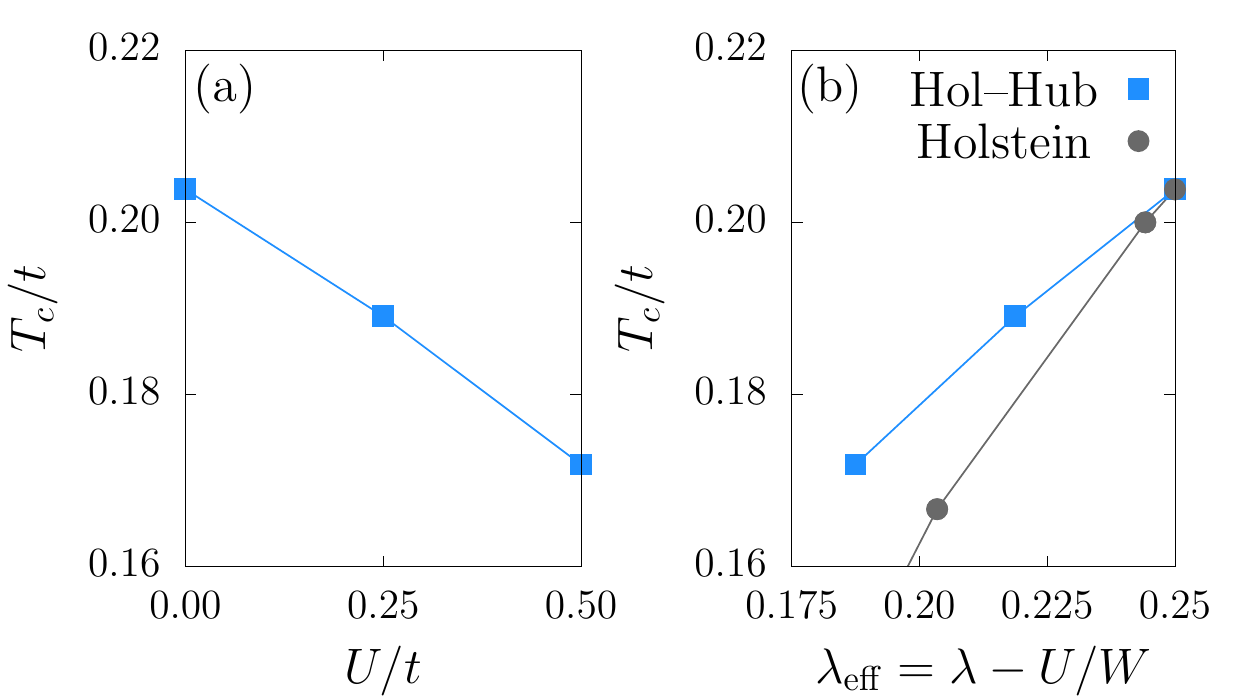}  
  \caption{\label{fig:TcvsU} Critical temperature of the CDW transition in
    the Holstein-Hubbard model. (a) Suppression of $T_c$ with increasing $U$
    at $\lambda=0.25$ from finite-size scaling, (b) comparison of Holstein
    and Holstein-Hubbard results in terms of the effective coupling
    $\lambda_\text{eff}=\lambda-U/W$. The points labeled `Holstein'
    correspond to $\lambda_c$ at different temperatures from Fig.~\ref{fig:phasediagram}. The points
    labeled `Hol-Hub' (Holstein-Hubbard) are for $T_c$ at $\lambda=0.25$ and
    $U/t=0,0.25,0.50$ from (a). Here, $\omega_0/t=0.1$.}
\end{figure}

The Hubbard repulsion suppresses CDW order
\cite{PhysRevB.52.4806,PhysRevLett.75.2570,PhysRevB.75.014503,PhysRevB.92.195102,PhysRevLett.109.246404,PhysRevB.87.235133,ohgoe2017competitions}.
This is already apparent from the effective Hubbard model~(\ref{eq:model2}) in the limit
$\omega_0\to\infty$ where a nonzero $U$ reduces the effective, attractive
interaction and thereby the CDW gap at $T=0$. Whereas CDW order is restricted
to $T=0$ in this limit, here we consider the Holstein-Hubbard model in the
opposite, adiabatic regime. Specifically, we take $\omega_0/t=0.1$ and $\lambda=0.25$.

To quantify the effect of $U$, we show in Fig.~\ref{fig:TcvsU}(a) the
suppression of $T_c$ as a function of $U$. Starting from
$T_c/t=0.204(1)$ at $U=0$, $T_c$ decreases by about 15 percent in the range
$U\in[0,0.5t]$. In principle, in the spirit of an effective Holstein model,
we can try to capture this effect by a coupling $\lambda_\text{eff}=\lambda-U/W$.
However, Fig.~\ref{fig:TcvsU}(b) reveals that for the parameters considered
this overestimates the effect of the Hubbard repulsion because $T_c$ at a
given $\lambda_\text{eff}$ in the Holstein model ($U=0$) is significantly
lower than in the Holstein-Hubbard model ($U>0$). We attribute this finding
to (i) the stronger suppression of the antiferromagnetic
correlations (long-range magnetic order only exists at $T=0$) compared to the
CDW correlations (CDW order exists also at $T>0$) at the temperatures considered,
and (ii) retardation effects. A DMFT analysis of the Holstein-Hubbard model revealed that $T_c$ is suppressed with increasing $U$
at weak electron-phonon coupling but initially enhanced at strong
coupling. This behavior was explained in terms of a reduction of the
bipolaron mass due to the onsite repulsion \cite{PhysRevLett.75.2570}.

\begin{figure}[b]
  \includegraphics[width=0.45\textwidth]{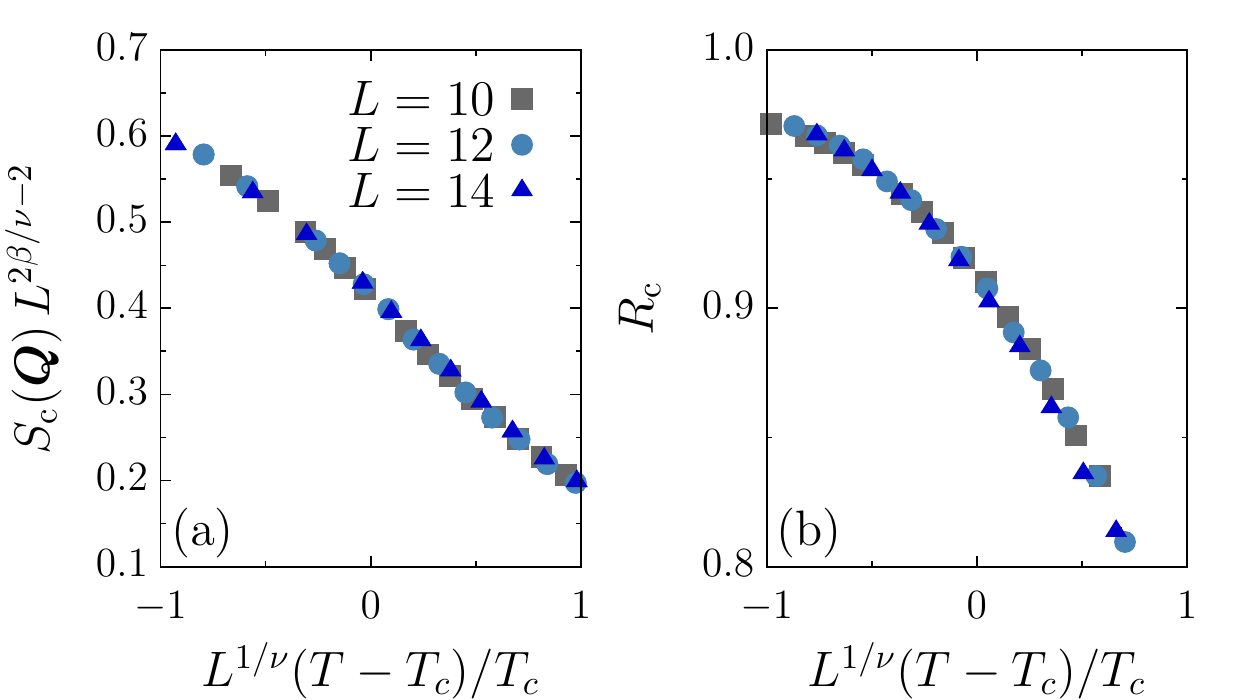}  
  \caption{\label{fig:quising} Scaling collapse of (a) the structure
    factor and (b) the correlation ratio for $\omega_0/t=0.1$,
    $\lambda=0.25$, and $U=0$ using the critical exponents of the 2D Ising
    model. The critical temperatures $T_c$ were determined from the best
    scaling collapse and are given in the text.
}
\end{figure}

\subsection{Critical behavior}\label{sec:results:critical}

In the thermodynamic limit, the long-range CDW
order at $T<T_c$ spontaneously breaks the sublattice symmetry. The two
possible CDW patterns (cf. Fig.~\ref{fig:phasediagram}) imply the same
critical behavior as the 2D Ising model and hence critical exponents
$\beta=1/8$ and $\nu=1$. Here, we demonstrate consistency with Ising
universality for $\om_0/t=0.1$ and $\lambda=0.25$.

As the order parameter, the charge structure
factor~(\ref{eq:scdw}) should obey the finite-size scaling relation
\cite{PhysRevLett.66.778} 
\begin{equation}\label{eq:scalingS}
  {S_\text{c}({\bm Q})}/{L^2}
 = L^{-2\beta/\nu}\, f_S[L^{1/\nu}(T-T_c)/T_c]\,.
\end{equation}
Therefore, plotting $S_\text{c}({\bm Q}) L^{2\beta/\nu-2}$ as a function of $L^{1/\nu}(T-T_c)/T_c$ 
should produce a collapse of the data onto the curve described by the scaling
function $f_S$. The best collapse \cite{autoscale}
over the interval $[-1,1]$ in Fig.~\ref{fig:quising}(a) gives $T_c/t=0.195(1)$, 
smaller than the value $T_c/t=0.204(1)$ (Fig.~\ref{fig:TcvsU}) determined from
finite-size scaling.

A similar analysis can be carried out for the correlation ratio, which is
expected to obey
\begin{equation}\label{eq:scalingR}
  R_\text{c} = f_R[L^{1/\nu}(T-T_c)/T_c]\,,
\end{equation}
involving only the correlation length exponent $\nu$. Hence, 
we expect a collapse onto $f_R$ by plotting $R_\text{c}$ as a function of
$L^{1/\nu}(T-T_c)/T_c$. The best collapse on $[-1,1]$ is
obtained for $T_c/t=0.205(1)$ and shown in Fig.~\ref{fig:quising}(b). This
critical value is consistent with the previous estimate  $T_c/t=0.204(1)$
in Fig.~\ref{fig:TcvsU}. However, the collapse exhibits stronger scattering
than for the structure factor, even though the correlation ratio is generally
expected to be less affected by finite-size corrections \cite{Binder1981}.

\subsection{Phase diagram}\label{sec:results:phasediagram}

Figure~\ref{fig:phasediagram} gives the finite-temperature phase diagram of
the Holstein model in terms of $T_c(\lambda)$, which separates the
low-temperature phase with long-range CDW order from the high-temperature
disordered phase. However, since accurate values of $T_c$ at very small
$\lambda$ are currently not accessible, Fig.~\ref{fig:phasediagram} does not
settle the question of whether or not the ground state has CDW order for any
$\lambda>0$. 

There are two well-understood limits. The {\it classical} Holstein
model ($\omega_0=0$) has a CDW ground state for any $\lambda>0$ and
$T_c>0$ (see Sec.~\ref{sec:results:Tc}). This follows from mean-field theory,
which becomes exact at $T=0$. In the opposite, antiadiabatic limit
$\omega_0\to\infty$, the Holstein model maps to the attractive Hubbard model,
whose ground state has coexisting CDW and superconducting order but $T_c=0$. 
Hence, as a function of $\omega_0$, the Holstein model interpolates between
two limits that both exhibit long-range CDW order at $T=0$. 

\begin{figure}[t]
  \includegraphics[width=0.25\textwidth]{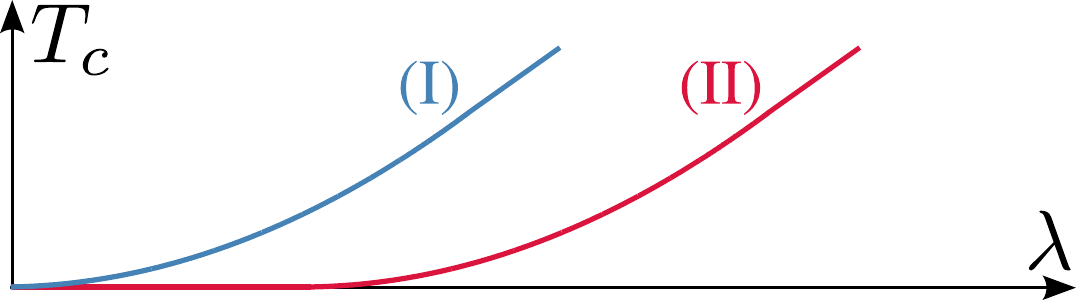}
  \caption{\label{fig:tcschematic} The two possible scenarios for the phase
    diagram of the Holstein model. In scenario (I), we have CDW order with
    $T_c>0$ for any $\lambda>0$. In scenario (II), $T_c=0$ for $\lambda<\lambda_c(\omega_0)$.}
\end{figure}

Between these limiting cases (\ie, for $0<\omega_0<\infty$), there appear to be 
two distinct scenarios for the shape of the phase boundary $T_c(\lambda)$, as
illustrated in Fig.~\ref{fig:tcschematic}. In
scenario (I), $T_c>0$ for any $\lambda>0$, so that the ground state is always
a CDW insulator. By contrast, in scenario (II), $T_c=0$ for
$\lambda<\lambda_c(\omega_0)$ and $T_c>0$ for $\lambda>\lambda_c(\omega_0)$. 
Case (II) can further be divided into (IIa) where CDW order exists at $T=0$
for any $\lambda$, and (IIb) with a disordered phase at $T=0$ below
$\lambda_c(\omega_0)$. In scenario (I), the adiabatic (classical) fixed
point determines the behavior for any finite $\omega_0$. On the other hand,
in scenario (IIa), the physics is determined by the antiadiabatic fixed point
for $\lambda<\lambda_c(\omega_0)$ and by the adiabatic fixed point for
$\lambda>\lambda_c(\omega_0)$. Note that CDW order with $T_c=0$ requires an
emergent continuous order parameter, as realized for the attractive Hubbard
model ($\omega_0=\infty$). However, the corresponding symmetry is broken for
$\omega_0<\infty$ by retardation effects in the Holstein model \cite{Hirsch83a}.

A CDW ground state for any $\lambda>0$ may be expected based on the
instability of the Fermi liquid. For the half-filled square lattice with
nearest-neighbor hopping, the noninteracting charge susceptibility
$\chi^{(0)}_\text{c}({\bm   Q})\sim\ln^2\beta t$ due to the
combined effect of nesting and Van Hove singularities
\cite{PhysRevLett.56.2732,PhysRevB.42.2416}. In the Hubbard model, such
divergences underlie the existence of an antiferromagnetic Mott insulator
for any $U>0$, and coexisting CDW and superconducting order for any $U<0$ \cite{Hirsch85}.
For the Holstein model that does not have a symmetry-imposed degeneracy of CDW and pairing
correlations, superconducting correlations were found 
to be weaker than CDW correlations at half-filling \cite{PhysRevB.42.2416},
consistent with the weaker divergence of the $\bm{Q}=0$ pairing susceptibility
$\chi^{(0)}_\text{p}({\bm Q})\sim\ln\beta t$.

Despite these theoretical arguments, metallic and superconducting ground
states were recently suggested for the half-filled Holstein and
Holstein-Hubbard models based on variational QMC simulations
\cite{ohgoe2017competitions,1709.00278}. A metallic phase is also found
within DMFT \cite{Koller04,PhysRevB.70.125114,PhysRevB.88.125126}, where a Van
Hove singularity is absent. For $\omega_0\ll t$, the results of
Fig.~\ref{fig:phasediagram} appear consistent with CDW
order even at $T=0$ for any $\lambda>0$. On the other hand, the phase
boundary $T_c(\lambda)$ in Fig.~\ref{fig:phasediagram} undergoes an
increasingly strong shift to larger $\lambda$ with increasing
$\omega_0$, in principle compatible with $T_c=0$ at sufficiently weak
coupling [scenario (II)].  In the significantly better understood 1D case,
numerical results show that for $\omega_0>0$ the ground state remains metallic
for $\lambda<\lambda_c$ despite a $\ln \beta t$ nesting-related
divergence of the charge susceptibility \cite{MHHF2017}. Since $T_c=0$ in the
1D case, this corresponds to scenario (IIb) above and is consistent with the
$\omega_0=\infty$ limit, the 1D attractive Hubbard model. The latter has a
metallic but spin-gapped Luther-Emery liquid \cite{Lu.Em.74} ground state and
no long-range order. Functional renormalization group calculations for the
2D Holstein-Hubbard model exclude metallic or superconducting behavior at
half-filling except for an extremely small region where $T_c$ is essentially
zero \cite{PhysRevB.92.195102}.

To address the ground-state phase diagram directly, we calculated the
correlation ratios
\begin{align}\label{eq:Rchic}
  R^\chi_\text{c} 
  &=  1-\frac{\chi_\text{c}(\bm{Q}-\delta{\bm
    q})}{\chi_\text{c}(\bm{Q})}\,,\quad
    \bm{Q}=(\pi,\pi)\,,
  \\
  R^\chi_\text{p} 
  &=  1-\frac{\chi_\text{c}(\bm{Q}-\delta{\bm q})}{\chi_\text{c}(\bm{Q})}\,,
    \quad \bm{Q} = (0,0)\,.
\end{align}
for CDW and s-wave pairing based on the susceptibilities
\begin{align}\label{eq:chic}
  \chi_\text{c}(\bm Q) 
  &= \frac{1}{L^2} \sum_{ij}
  e^{\rmi(\bm{r}_i-\bm{r}_j)\cdot\bm{Q}} \int_0^\beta \rmd \tau
  \las \on_{i}(\tau) \on_{j}\ras\,,
  \\\label{eq:chip}
  \chi_\text{p}(\bm Q) 
  &= \frac{1}{L^2} \sum_{ij}
  e^{\rmi(\bm{r}_i-\bm{r}_j)\cdot\bm{Q}} \int_0^\beta \rmd \tau
  \las \hat{\Delta}^\dag_{i}(\tau) \hat{\Delta}^\nag_{j}\ras\,,
\end{align}
where $\hat{\Delta}_i=c_{i\UP}c_{i\DO}$. The susceptibilities generally
exhibit better finite-size scaling behavior than the corresponding static
structure factors [cf. Eq.~(\ref{eq:scdw})]. We take a coupling
$\lambda=0.075$, for which Refs.~\cite{ohgoe2017competitions,1709.00278}
suggest the absence of CDW order at $U=0$ over a large range of phonon
frequencies. The inverse temperature was scaled as $\beta t=2L$ (with $4\leq
L\leq 16$), which is at the current limit of the CT-INT method due to the sign problem. 

\begin{figure}[t]
  \includegraphics[width=0.45\textwidth]{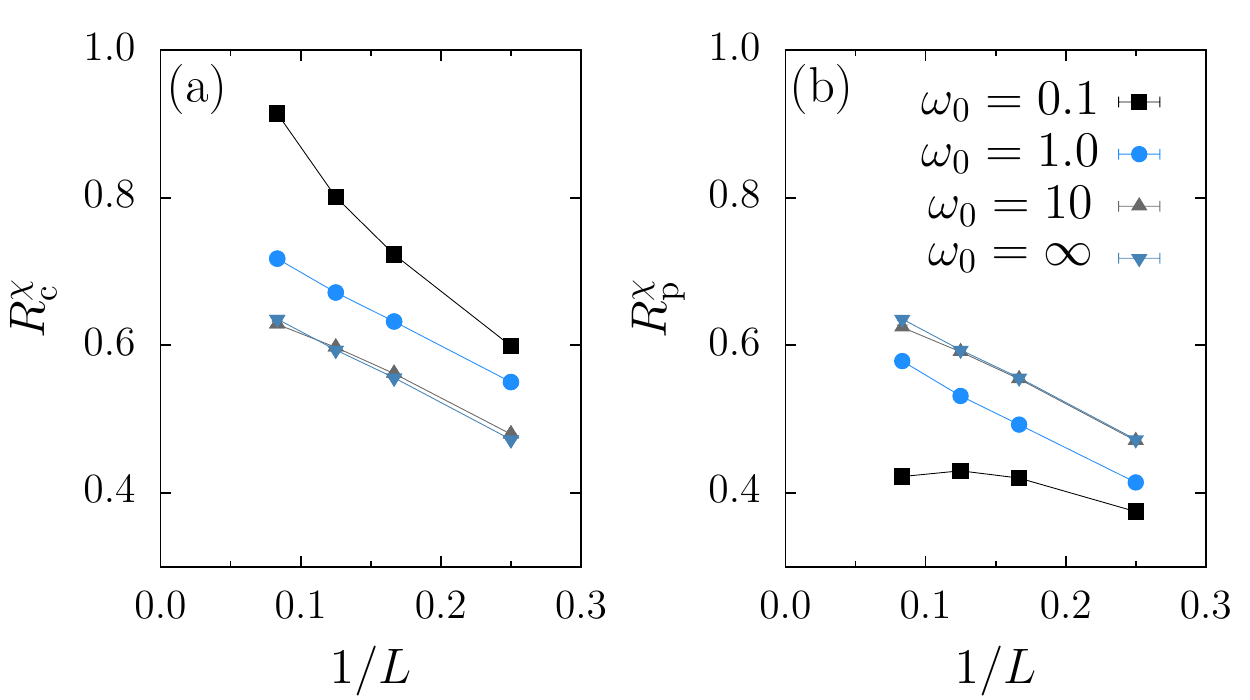}
  \caption{\label{fig:beta2L} (a) Charge and (b) pairing correlation ratios
    for different phonon frequencies. Here, $\beta t=2L$, $\lambda=0.075$,
    $U=0$.}
\end{figure}

\begin{figure}[t]
  \includegraphics[width=0.45\textwidth]{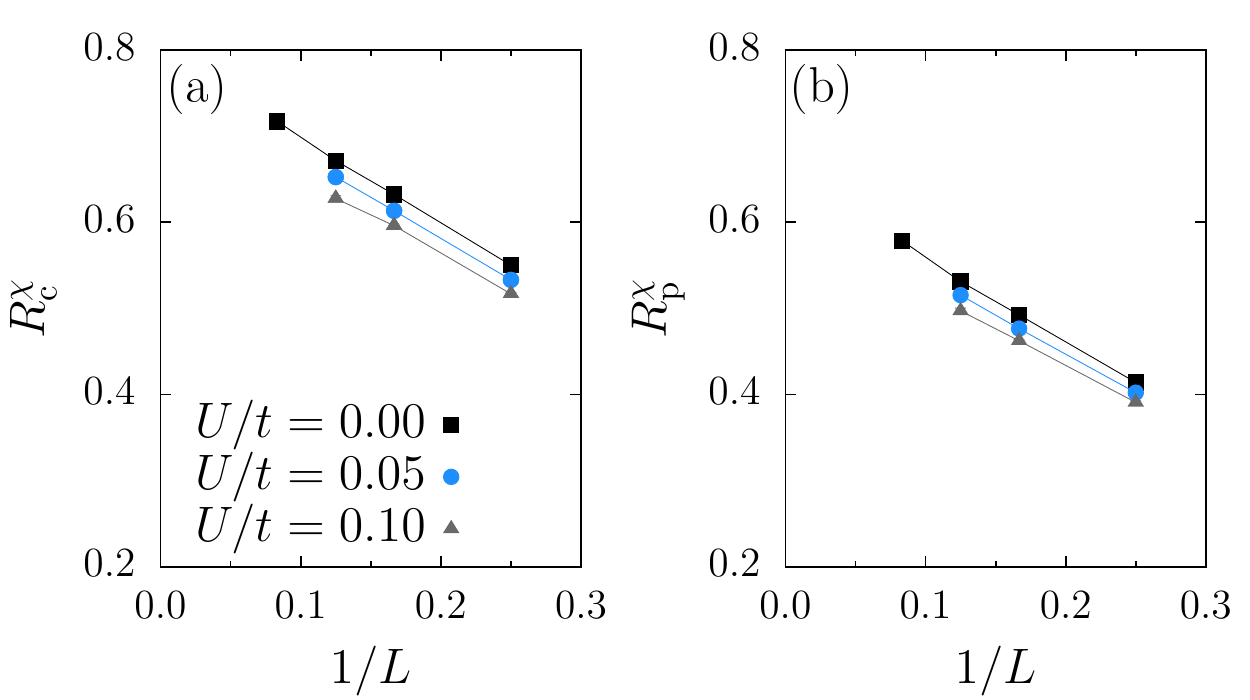}    
  \caption{\label{fig:beta2L_U} (a) Charge and (b) pairing correlation ratios
    for different Hubbard repulsions. Here, $\beta t=2L$, $\lambda=0.075$,
    $\omega_0/t=1$.}
\end{figure}

The correlation ratios shown in Figs.~\ref{fig:beta2L} and~\ref{fig:beta2L_U} 
have the same properties as discussed in Sec.~\ref{sec:results:Tc};
long-range order is revealed by $R^\chi_\alpha\to 1$ for $L\to\infty$, and
a larger correlation ratio indicates stronger correlations in the
corresponding channel. For $\om_0=0.1$, the results in Fig.~\ref{fig:beta2L}(a)
suggest long-range CDW order, consistent with Fig.~\ref{fig:phasediagram}.
At the same time, the pairing correlation ratio in Fig.~\ref{fig:beta2L}(b)
is strongly suppressed. Upon increasing $\omega_0$, CDW correlations are
suppressed and pairing correlations enhanced, but
$R^\chi_\text{c}>R^\chi_\text{p}$ for any $\omega_0<\infty$. Degenerate CDW
and pairing correlations are only observed for the attractive Hubbard model
($\omega_0=\infty$). 
The fact that CDW correlations at $\omega_0<\infty$ are stronger
than for $\omega_0=\infty$ suggests a CDW ground state also for the Holstein
model and likely no superconducting order since $T_c$ is already minimal for
$\omega_0=\infty$. As demonstrated in Fig.~\ref{fig:beta2L_U}, a nonzero
Hubbard repulsion suppresses both CDW and pairing correlations while
enhancing antiferromagnetic correlations (not shown).

Figure~\ref{fig:beta2L} also reveals that in the weak-coupling regime where
an absence of CDW order was predicted \cite{ohgoe2017competitions,1709.00278},
it is challenging to unequivocally detect the known $T=0$ long-range
order of the attractive Hubbard model in terms of $R^\chi_\text{c},R^\chi_\text{p}\to 1$ for
$L\to\infty$. The same should be true for the Holstein and Holstein-Hubbard
model in the regime where $T_c$ is small. Therefore, leaving aside the approximations inherent to
variational QMC methods, the reported absence of CDW order 
\cite{ohgoe2017competitions,1709.00278} should also be taken with care.

While we are unable to provide a definitive $T=0$ phase diagram, the
results of Fig.~\ref{fig:beta2L} together with the
observation that long-range CDW order is known to exist at $T=0$ for both
$\om_0=0$ and $\om_0=\infty$ are consistent with CDW order but no
superconductivity in the half-filled Holstein model at $T=0$. 
Furthermore, in the absence of a higher symmetry relating CDW and
superconducting order as in the attractive Hubbard model, we expect $T_c>0$
(although potentially exponentially small) and hence
scenario (I) depicted in Fig.~\ref{fig:tcschematic}.

\subsection{Bipolaron liquid}\label{sec:results:bipolarons}

\begin{figure}[t]
  \includegraphics[width=0.45\textwidth]{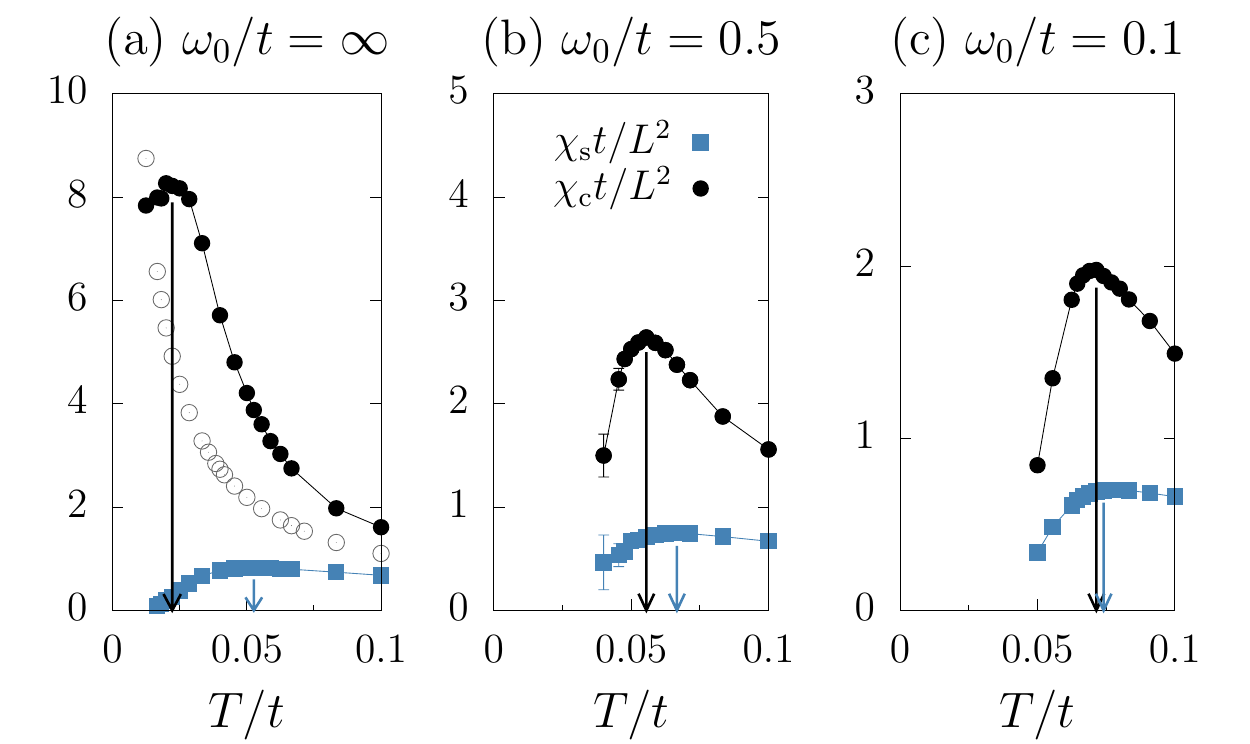}
  \caption{\label{fig:atthubbard2} Local spin and charge susceptibilities
    [Eq.~(\ref{eq:localsusc})] for  $\lambda=0.1$, $U=0$, and $L=8$. Open
  symbols in (a) are for $\lambda=0$,
    arrows indicate maxima.}
\end{figure}

A final interesting point is the
nature of the metallic phase at $T>T_c$. In the CDW phase, spin, charge, and
hence also single-particle excitations are gapped. For 1D electron-phonon models, the
spin gap persists in the metallic phase \cite{MHHF2017} and the $T=0$ CDW
transition occurs at the two-particle level via the ordering of preformed
pairs (singlet bipolarons) and the opening of a charge gap. The same is true for
the 2D attractive Hubbard model for which the spin gap can be made
arbitrarily large by increasing $U$ while keeping $T_c=0$. Hence, the
disordered phase at low but finite temperatures is not a Fermi liquid but a metal with gapped
single-particle and spin excitations \cite{PhysRevB.50.635,PhysRevLett.69.2001}, the 2D analog of a Luther-Emery liquid \cite{Lu.Em.74}. Singlet bipolarons
in principle also form for any $\lambda>0$ in the 2D Holstein model, 
although their binding energy ($\sim \lambda$) can be small
\cite{PhysRevB.69.245111}. Nevertheless, we expect a spin-gapped metallic
phase for suitable parameters. At
sufficiently high temperatures, bipolarons undergo thermal
dissociation \cite{PhysRevB.71.184309}.

To detect signatures of a spin-gapped metal, we consider the static charge and spin susceptibilities
\begin{equation}\label{eq:localsusc}
\chi_\text{c} =\beta (\las \hat{N}^2\ras - \las\hat{N}\ras^2), \quad 
\chi_\text{s} =\beta (\las \hat{M}^2\ras - \las\hat{M}\ras^2)\quad 
\end{equation}
with $\hat{N} = \sum_i \hat{n}_i$, $\hat{M} = \sum_i \hat{S}^x_i$.
Figure~\ref{fig:atthubbard2}(a) shows results for $\lambda=0.1$ and
$\omega_0/t=\infty$. Whereas $\chi_\text{s}/L^2$ diverges with decreasing
temperature in a Fermi liquid (open symbols), it is strongly suppressed 
as $T\to0$ by the spin gap. The charge susceptibility is also suppressed at
very low $T$, but $\chi_\text{c}/L^2$ approaches a finite value determined by
the density of $T=0$ charge fluctuations. The distinct temperature scales reflected by the maxima of $\chi_\text{s}/L^2$ and
$\chi_\text{c}/L^2$ reveal the spin-gapped metallic phase at $T>0$ in the
attractive Hubbard model. For the
Holstein model, $\chi_\text{s}/L^2$ is cut off by the spin
gap, whereas $\chi_\text{c}/L^2$ is cut off by the charge gap that
appears at the CDW transition at $T=T_c$. The distinct maxima visible even in the
adiabatic regime [Figs.~\ref{fig:atthubbard2}(b)
and~\ref{fig:atthubbard2}(c)] are consistent with a spin-gapped phase at
$T>T_c$. The extent of the latter appears to decrease with
decreasing $\omega_0/t$ and the phase is expected to be absent
in the classical or mean-field limit ($\omega_0=0$) where charge and spin gaps become
equal. An immediate and important corollary of the existence of a spin-gapped
metal of bipolarons above $T_c$ would be that, contrary to expectations in previous work
\cite{PhysRevB.48.7643,PhysRevB.48.16011}, the appearance
of a gap in the density of states does in general not imply CDW order.
The additional spin-gap component is also compatible with
experimentally observed large gap to $T_c$ ratios \cite{PhysRevB.63.115114}.

In principle, a spin-gapped phase without long-range order (CDW or
superconductivity) could also exist at $T=0$, but the discussion in
Sec.~\ref{sec:results:phasediagram} provided arguments against a disordered phase.
While well established in 1D electron-phonon models in terms of a Luther-Emery
liquid \cite{MHHF2017}, it would correspond to a so-called Bose metal
\cite{PhysRevB.60.1261} in higher dimensions. An interesting question
regarding the recent findings of
Refs.~\cite{ohgoe2017competitions,1709.00278} is whether the variational wave
functions used can distinguish between spin-gap formation and
superconductivity. To this end, it would be useful to test this method for
the intricate but well understood 1D Holstein model.

\section{Conclusions}\label{sec:conclusions}

We applied exact, continuous-time QMC simulations to the half-filled
Holstein-Hubbard model on the square lattice. The critical temperature for
the CDW transition was determined as a function of phonon frequency,
electron-phonon coupling, and Hubbard repulsion from finite-size scaling. We
also demonstrated the expected 2D Ising universality of this transition and
addressed the ground-state phase diagram, providing data and theoretical
arguments for the likely absence of a metallic or superconducting phase at
weak coupling. Finally, we discussed the possibility of a spin-gapped metallic
phase of bipolarons above $T_c$. The quantitative ground-state phase diagram
remains an  important open problem.

\vspace*{1em}
\begin{acknowledgments}
We thank F. Assaad, N. Costa, P. Br\"ocker, T. Lang, and R. Scalettar for helpful discussions and the DFG for
support via SFB 1170 and FOR~1807. We gratefully acknowledge the
computing time granted by the John von Neumann Institute for Computing (NIC)
and provided on the supercomputer JURECA \cite{jureca} at the J\"{u}lich
Supercomputing Centre.
\end{acknowledgments}

%\bibliography{../refs}


\end{document}